# Nuclear Analysis and Shielding Optimisation in Support of the ITER In-Vessel Viewing System Design

Andrew Turner[a], Raul Pampin[b], M.J. Loughlin[c], Zamir Ghani[a], Gemma Hurst[a], Alessandro Lo Bue[b], Samuel Mangham[a], Adrian Puiu[b], Shanliang Zheng[a]

[a] *EURATOM/CCFE Fusion Association, Culham Science Centre, Abingdon, Oxon, OX14 3DB, UK*
[b] *F4E Fusion for Energy, Josep Pla 2, Torres Diagonal Litoral B3, 08019 Barcelona, Spain*
[c] *ITER Organisation, Route de Vinon sur Verdon, 13115 Saint Paul Lez Durance, France*

The In-Vessel Viewing System (IVVS) units proposed for ITER are deployed to perform in-vessel examination. During plasma operations, the IVVS is located beyond the vacuum vessel, with shielding blocks envisaged to protect components from neutron damage and reduce shutdown dose rate (SDR) levels. Analyses were conducted to determine the effectiveness of several shielding configurations. The neutron response of the system was assessed using global variance reduction techniques and a surface source, and shutdown dose rate calculations were undertaken using MCR2S.

Unshielded, the absorbed dose to piezoelectric motors (PZT) was found to be below stable limits, however activation of the primary closure plate (PCP) was prohibitively high. A scenario with shielding blocks at probe level showed significantly reduced PCP contact dose rate, however still marginally exceeded port cell requirements. The addition of shielding blocks at the bioshield plug demonstrated PCP contact dose rates below project requirements. SDR levels in contact with the isolated IVVS cartridge were found to marginally exceed the hands-on maintenance limit. For engineering feasibility, shielding blocks at bioshield level are to be avoided, however the port cell SDR field requires further consideration. In addition, alternative low-activation steels are being considered for the IVVS cartridge.

## 1. Introduction

The In-Vessel Viewing System (IVVS) proposed for ITER consists of six identical units, which are deployed between pulses or during shutdown to perform in-vessel inspections of plasma facing components, and estimate in-vessel dust quantities [1]. During plasma operation, the system is housed in dedicated ports at B1 level (Fig. 1), with deployment between the divertor cassettes and the lowermost outboard blanket modules (Fig. 2).

In order to progress the design of the IVVS, nuclear analysis was performed by Culham Centre for Fusion Energy (CCFE) on behalf of Fusion for Energy. Boron carbide shielding blocks are envisaged to protect sensitive components of the IVVS during plasma operations, and to minimise the dose to maintenance personnel due to activation of the IVVS system and the primary closure plate (PCP). Analyses were conducted using MCNP [2] to determine the acceptability of several shielding configurations.

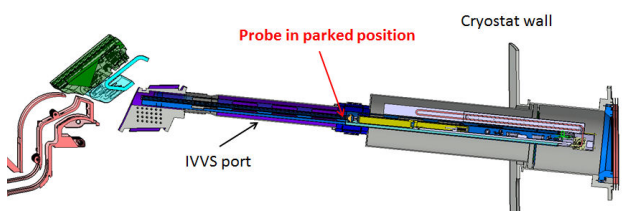

Fig. 1. IVVS in parked position

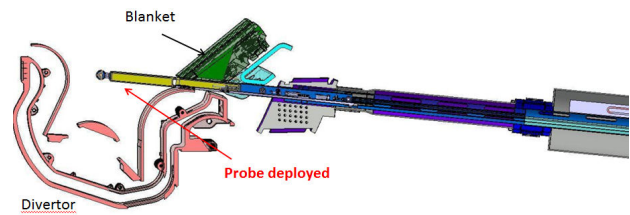

Fig. 2. IVVS in-vessel deployment

## 2. IVVS Neutronics Model

### 2.1. CAD simplification

A CAD model of the IVVS was simplified and prepared for neutronics analysis. Small features unnecessary for neutron transport were removed using SpaceClaim®. The resulting CAD model is shown along with key features, in Fig. 3. Conversion of the solid and void bodies to MCNP geometry was performed using MCAM [3], and materials assigned. Finally, the MCNP model was 'cleaned' from minute geometry imperfections intrinsic to the conversion process. The clean IVVS MNCP model was integrated into the ITER reference MCNP model B-lite V2.

_________________________________________________________________________

author's email: andrew.turner@ccfe.ac.uk

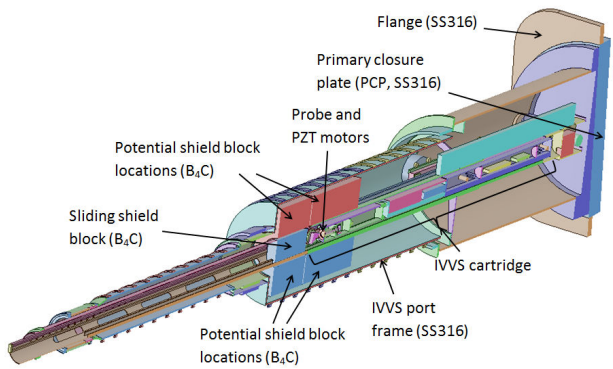

Fig. 3. Cut-away model of the IVVS port and cartridge

## 2.2. Integration with ITER reference model

Reference MCNP models of the ITER device, constructed as 40° sectors of the tokamak, are provided by the ITER Organisation (IO) for use in analysis tasks. The B-lite V2 model [4] was specified for this analysis. The B-lite model consists of universes, permitting users to swap individual component models within a larger geometrical framework. A universe container cell for the IVVS did not exist in the standard model, hence a bounding geometry was created and integrated into the existing top level universe cell structure via the use of the complement (#) operator. This bounding universe was also used to produce a penetration in the triangular support and blanket. A representation of the IVVS vacuum vessel penetration was similarly included.

The location of the integrated IVVS universe is between two lower ports. In the standard B-lite model, this would be close to the model boundary as shown in Fig. 4. This was not desirable due to the presence of reflective boundary conditions; additionally a central position is preferred in order to simplify the use of mesh tallies. The existing lower port universe was modified to permit a central placement of the IVVS whilst still being faithful to the relative position of the surrounding ports and coils as shown in Fig. 4.

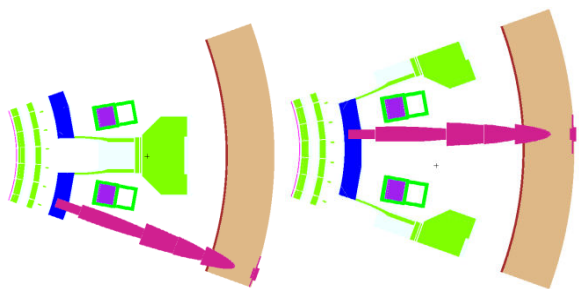

Fig. 4. B-lite MCNP model showing lower ports and IVVS universe (left=standard, right=modified lower ports)

Further modifications to the B-lite V2 reference model included a more realistic bioshield plug representation, a 'generic' diagnostics equatorial port plug, and the inclusion of a model of the cryopump in the lower ports. The integrated model is shown in Fig. 5.

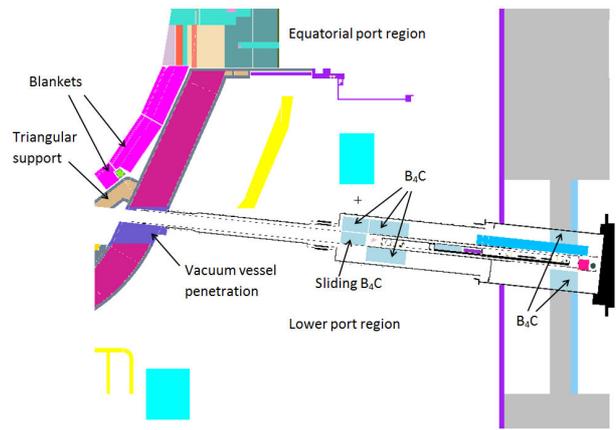

Fig. 5. IVVS model integrated in B-lite (scenario S2 shown)

## 2.3. Shielding scenarios

MCNP models were produced for the following shielding configurations, shown in Fig. 6:

- S0: the empty IVVS port, with no internal components or shielding, to provide 'baseline' results.
- S1: the IVVS port and internal components, plus boron carbide shield blocks protecting the IVVS probe area.
- S2: as S1, with the addition of boron carbide shield blocks in the vicinity of the bioshield, to reduce dose rates in the port cell where restrictions apply.
- S2*: as S2, with the removal of the sliding shield block, to assess the impact of a scenario where the shield had failed to return to the correct position.

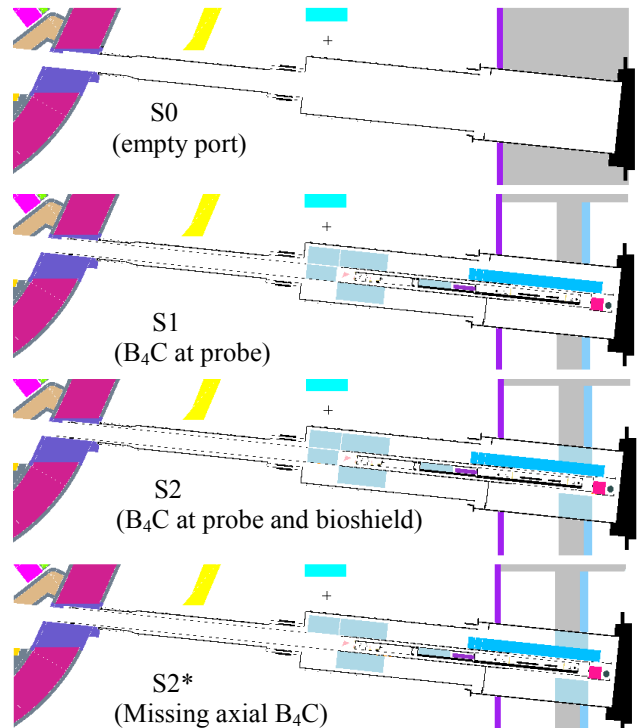

Fig. 6. IVVS shield block configurations S0, S1, S2, S2*

## 3. Tallies and Methodology

### 3.1. Tallies

Calculations carried out using MCNP5 and FISPACT-II [5] determined:

- Neutron flux spectrum, material damage, helium production and absorbed dose rates in the piezoelectric motors, to confirm reliable operation. Damage and helium production rates were determined using FISPACT-II and TENDL [6] libraries, in preference to MCNP tally multipliers. Previous experience has shown that DPA and gas production results obtained using FISPACT-II tend to be higher (and therefore more conservative) than those obtained using MCNP.

- Neutron flux spectrum and contact dose rate in the PCP. Access to the port cell region immediately behind the PCP is required for maintenance, and port cell project requirements dictate a maximum dose rate of 10 $\mu$Sv/hr at 24 hours after operations, at 30 cm from a surface. The contact dose rate produced by FISPACT-II was used to be conservative, and avoid a computationally expensive 3-D activation calculation.

- Material damage rate and helium production in the vacuum vessel wall and divertor pipes. Penetrations in the vacuum vessel, blanket modules and triangular support permit the IVVS to enter the plasma chamber, however these penetrations reduce the shielding and thus the local increase in radiation load was to be examined.

- Nuclear heating and fast neutron flux to superconducting coils. The IVVS vessel penetration must not lead to a significant local increase in heat deposition in the ITER coils.

- Absorbed dose rate to PZT motors after shutdown, and biological dose rate due to activation of the IVVS removable cartridge. A section of the IVVS is intended to be removed and to undergo hands-on maintenance $10^6$ seconds after machine shutdown, for which a 100 $\mu$Sv/hr limit applies at the location maintenance will be conducted. The MCR2S code [7] was used to determine the activation gamma source using the SA2 irradiation scenario [8] for required decay times. In order to calculate the shutdown absorbed dose field for the parked IVVS, the activation of large regions of the ITER device were simulated; due to memory constraints, the activation source was generated on a 6 cm resolution mesh local to the IVVS, and at a 15 cm resolution in neighbouring regions.

### 3.2. Global Variance Reduction

Global variance reduction (GVR) [9] is required to accurately compute the tally contribution at the IVVS from neutrons penetrating shielded regions of ITER (such as the blankets and vacuum vessel). CCFE has previously developed a method of implementing weight windows to perform GVR and obtain a result with near-uniform statistical accuracy in all problem space (a 'flattening' of the variance). This is done through a series of iterations using the forward flux solution to produce a weight window map. Since the use of GVR is computationally demanding, it was decided to produce a weight window mesh that covered only part of the B-lite geometry that was presumed to be the most important to the IVVS region. By allowing particles outside of this region to be transported in analogue mode, computational time would not be wasted obtaining accurate transport in regions where deep penetration effects are unimportant, such as the inboard side.

### 3.3. Surface source

Initial calculations showed that neutrons streaming down the IVVS penetration had a significant effect on the system nuclear responses. To obtain statistically accurate results, it was necessary to run a large number of histories to adequately sample the vessel penetration, which proved to be impractical under GVR.

In order to resolve these two conflicting requirements of high source sampling and deep penetration, the calculation was split into two runs, one using the surface source feature of MCNP to sample the streaming contribution, and the other using GVR but with any particles streaming down the IVVS 'killed' to prevent any double counting. The results of the two calculations were then summed, and the statistical errors combined in quadrature. In this way, the separate contributions could be assessed to higher statistical accuracy than would have otherwise been possible; in addition this method has the advantage that the dominant source (streaming or global effects) for the tally of interest is provided.

## 4. Results and Discussion

The neutron fluxes at the primary closure plate (PCP) are given in Table 1 and for the piezoelectric (PZT) motors in Table 2. Table 3 shows material damage rates and helium production in the PZT actuators. Results are normalised to 500 MW fusion power, and where applicable, integrated over the lifetime of ITER ($1.7 \times 10^7$ s). Some values were not explicitly calculated within MCNP, for example, total absorbed dose in neutron-only calculations. These were instead estimated based on the ratio of neutron flux and indicated as (est). Contact dose, damage and helium production rates were determined using FISPACT-II, and thus no statistical errors are available.

Table 1. PCP neutron flux and contact dose rate

| Scenario | Neutron flux (n/cm$^2$/s) | Contact dose rate at 24 hr ($\mu$Sv/hr) |
|---|---|---|
| S0 | $8.8 \times 10^9$ (6%) | $1.4 \times 10^4$ |
| S1 | $1.5 \times 10^7$ (3%) | 19.6 |
| S2 | $5.2 \times 10^5$ (9%) | 4.6 |
| S2* | $2.9 \times 10^6$ (30%) | 150 (est) |

Table 2. PZT flux and absorbed dose

| Scenario | Neutron flux (n/cm$^2$/s) | EoL absorbed dose (Gy) |
|---|---|---|
| S0 | 2x10$^{10}$ (est) | 2x10$^6$ (est) |
| S1 | 4.3x10$^8$ (20%) | 5x10$^4$ (est) |
| S2 | 3.4x10$^8$ (15%) | 4.07x10$^4$ (11%) |
| S2* | 1.4x10$^{10}$ (8%) | 2x10$^6$ (est) |

Table 3. PZT damage and helium production

| Scenario | EoL damage (DPA) | EoL helium (appm) |
|---|---|---|
| S0 | N/C | N/C |
| S1 | 7.3x10$^{-6}$ | 9.5x10$^{-6}$ |
| S2 | 4.7x10$^{-6}$ | 9.4x10$^{-6}$ |
| S2* | 2.0x10$^{-4}$ | 4.1x10$^{-4}$ |

*est=estimated, N/C=not computed, EoL = end of life.*

Results for the unshielded configuration, S0, revealed that neutrons streaming through the penetration dominates the IVVS nuclear responses. Piezoceramic materials have been proven stable up to doses in excess of 4 MGy [10], and unshielded the lifetime absorbed dose is 2 MGy, below proven stable levels. However, the PCP contact dose rate is prohibitively high compared to the project requirement of 10 μSv/hr, hence shielding is required to attenuate the streaming neutrons and reduce the dose rate in the port cell.

The S1 scenario included shielding blocks at probe level, and results showed the PCP contact dose rate still marginally exceeded the requirements. To further reduce dose levels, additional shielding blocks were modelled at the level of the bioshield (configuration S2), the results of which showed PCP contact dose rate below 10 μSv/hr. The S2* configuration showed unacceptable levels of PCP activation in the event of a failure scenario.

Results of the more comprehensive analysis conducted for scenario S2 provided additional data. The nuclear heating during plasma operation in the axial shield block is ~1 W, 80% of which is deposited in the first 10 cm; very low heating was found in other shield blocks and IVVS components. During shutdown, the activation of the IVVS and surrounding ITER components gives rise to a dose field, however the resulting absorbed dose and heating rates to IVVS components were found to be several orders of magnitude lower than those during a plasma pulse, and not significant.

Biological dose rates were determined for an isolated IVVS cartridge, shown in Fig. 7. This analysis included only the activation of the cartridge within the IVVS (port shown for visual reference, though not activated). Table 4 shows dose rates extracted along a line radially from the centre of the cartridge (also shown in Fig. 7).

Table 4. Shutdown dose rate (μSv/hr) for scenario S2

| Distance | Dose rate (μSv/hr) | | |
|---|---|---|---|
| Time | 10$^4$ s | 10$^5$ s | 10$^6$ s |
| Contact | 580 | 160 | 140 |
| 20 cm | 170 | 43 | 39 |
| 100 cm | 35 | 8.8 | 8.0 |

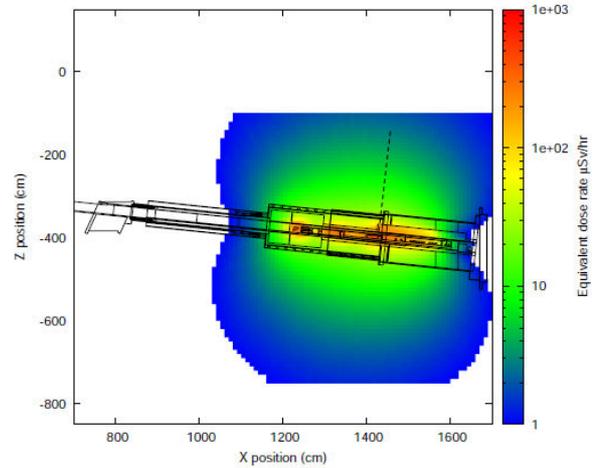

Fig. 7. SDR due to activation of IVVS cartridge (t=10$^5$ s)

The SDR in contact with the isolated IVVS cartridge was 140 μSv/hr at 10$^6$ s, marginally exceeding the hands-on maintenance limit of 100 μSv/hr. This dose rate drops rapidly with distance, and is 39 μSv/hr at 20 cm. The uncertainty on this result is difficult to define, with a systematic error due to mesh volume averaging of neutron fluxes and materials, as well as a statistical error associated with the neutron flux spectrum. A rigorous treatment of uncertainty propagation is a topic of current research, however a study on simplified geometry suggests the uncertainty on dose rate may be up to twice that of the total neutron flux uncertainty (errors on the total neutron flux were less than 10%). Deviations due to mesh resolution are estimated at 25% [11], and thus the total uncertainty is approximately 50%.

The peak neutron flux, material damage rate and helium production rate were assessed for the divertor cooling pipes and vacuum vessel immediately behind the cut-out in the blanket and triangular support (Table 5). The cut-out was found to increase damage rates and helium production in the divertor pipe region and vacuum vessel by a factor of 3 immediately behind the cut-out, when compared with values obtained with no such cut-out. The values are still within ITER project limits. Additionally it was demonstrated that the IVVS vessel penetration has a negligible effect on magnetic coil nuclear loads, with a total increase in TF coil nuclear heating of 0.8% for six such IVV systems.

Table 5. Effect of the IVVS cut-out

| Location | Flux (n/cm$^2$/s) | EoL damage (DPA) | EoL Helium (appm) |
|---|---|---|---|
| Divertor pipe | 3.17x10$^{12}$ | 1.30x10$^{-2}$ | 2.20x10$^{-1}$ |
| Vacuum vessel | 3.11 x10$^{12}$ | 9.22x10$^{-3}$ | 1.92x10$^{-1}$ |

## 5. Summary and Conclusions

Analyses were conducted using MCNP to determine the acceptability of the current IVVS design for four shielding configurations. Results were obtained for the neutron flux, nuclear heating and absorbed dose rates, helium production and material damage rates, as well as for the neutron flux and spectrum in the closure plate. The SDR field was assessed for scenario S2, both for activation of the IVVS and surrounding ITER components and for the removable IVVS cartridge. The effect of the IVVS penetration on the divertor pipes, vacuum vessel and magnetic coils were determined. Analysis on the IVVS design has concluded that:

- In an unshielded configuration, the end-of-life (EoL) absorbed dose to PZT is below stable limits. The PCP contact dose rate is prohibitively high.

- S1 scenario showed significantly reduced PCP contact dose rate, however this still exceeded the 10 µSv/hr port cell project requirements by a factor of 2.

- The presence of additional shielding blocks at the level of the bioshield (configuration S2), demonstrated PCP dose rates below 10 µSv/hr.

- The S2* configuration, in which the movable shield block failed to return to the correct position, showed high levels of PCP activation.

- Absorbed dose and heating rates at shutdown to the PZT are several orders of magnitude lower than those during a plasma pulse and are not significant.

- The effect of the IVVS cut-out in the blanket and triangular support is to increase damage rates and helium production in the immediate vicinity by a factor of 3, however the values are within project limits. The IVVS vessel penetration has a negligible effect on magnetic coil nuclear loads.

- Nuclear heating rates in the shielding blocks and other IVVS components are very small.

- Shutdown dose rates in contact with the isolated IVVS cartridge are in the range 100-140 µSv/hr at $t=10^6$ seconds, exceeding hands-on maintenance limits of 100 µSv/hr.

The modelling and calculations reported here assisted the pre-conceptual engineering design of the IVVS for ITER. The surface source approach greatly improved analysis turnover and statistics, and provided valuable information on the relative magnitude of these components for the different scenarios analysed.

It should be noted that while comparisons with dose limits were undertaken at this stage of the design process, this is not a sufficient condition for design acceptability and future studies must also show dose rates are as low as reasonably achievable (ALARA).

The S1 scenario is preferred over S2 for engineering feasibility, and these analyses have shown PCP contact dose for S1 is only marginally above port cell project requirements. Detailed analysis of the radiation fields in the port cell area are being performed for the next stage of the design process, and the use of alternative low-activation steels are under consideration.

## Acknowledgments

This work was funded by F4E under contract F4E-2008-OPE-002-01-06. The views and opinions herein do not necessarily reflect those of F4E, which is not liable for any use that may be made of the information contained herein.

This work was carried out using an adaptation of the B-lite MCNP model which was developed as a collaborative effort between the FDS team of ASIPP China, University of Wisconsin-Madison, ENEA Frascati, CCFE UK, JAEA Naka, and the ITER Organisation.